\documentclass{article}
\usepackage{listings}
\usepackage{amsmath}

\newcommand{\rr}{r\nobreakdash}	% for "\r-wartość"

\usepackage[utf8x]{inputenc}
\usepackage{polski}
\usepackage[polish]{babel}
\usepackage{listings}
\usepackage{color}

\usepackage[breaklinks,backref]{hyperref}
\hypersetup{colorlinks=true,linkcolor=black,citecolor=black,filecolor=black,urlcolor=black,breaklinks=true}

\title{C++11 -- idea r-wartości i~przenoszenia}
\author{Piotr Beling \\ Uniwersytet Łódzki, Wydział Matematyki i Informatyki}
\date{\today}

\begin{document}
\maketitle

\begin{abstract}
Niniejszy artykuł jest jednym z~serii artykułów w~których zawarto przegląd nowych elementów języka C++ wprowadzonych przez standard ISO/IEC 14882:2011, znany pod~nazwą C++11.

W~artykule przedstawiono nowe możliwości związane z~przekazywaniem parametrów i~pisaniem konstruktorów. Zawarto w nim dokładne omówienie idei \rr-wartości i~przenoszenia obiektów.
\end{abstract}
% This paper presents a review of some new futures introduced to C++ language by ISO/IEC 14882:2011 standard (known as C++11).
% It describes the ideas of r-values and move constructors.

\tableofcontents

\cleardoublepage

\section{Wstęp}
Standard języka C++, znany jako C++11 (a także C++0x), został opublikowany we~wrześniu 2011 roku, w~dokumencie ISO/IEC 14882:2011\footnote{Dokument ten dostępny jest odpłatnie, jednakże ze strony \url{http://www.open-std.org} można pobrać jego darmową wersję roboczą \cite{CPP11STD}. Najnowsza (N3337) datowana jest na 16 stycznia 2012.}.

W~stosunku do~poprzedniego standardu (C++03), wprowadza on dużą liczbą nowych elementów, zarówno do samego języka, jak i do~jego standardowych bibliotek.

W~większość popularnych kompilatorów dość szybko zaimplementowano prawie wszystkie nowe elementy wprowadzone w~C++11.

Niniejszy artykuł, jeden z serii, stanowi krótki i~bogato zilustrowany przykładami przegląd nowych konstrukcji które wzbogaciły język.

Szczególną uwagę przywiązano w~nim do możliwość związanych z~przekazywaniem parametrów i pisaniem konstruktorów. Szczególnie dokładne omówiona jest idea \rr-wartości i~przenoszenia wartości obiektów.

\section{R-wartości}
Terminem \rr-wartość (ang. \rr-value) określa się wartości mogące wystąpić jedynie po prawej stronie operatora przypisania, takie jak obiekty tymczasowe. Można je też postrzegać jako zmienne które nie będą dalej używane, aż do~momentu ich~skasowania, np. we~fragmencie programu:
\begin{lstlisting}
z.setValue(Obj());
\end{lstlisting}
obiekt tymczasowy \lstinline|Obj| jest \rr-wartością. Zostanie on~skasowany w~miejscu wystąpienia średnika na~końcu linii i w~związku z~tym nie będzie używany po~powrocie z~metody \lstinline|setValue|.
Proszę zauważyć, że~ostatnia informacja może być bardzo cenna dla~możliwości efektywnego zaimplementowania tej~metody (tymczasem w~C++03 nie możliwości jej~wygodnego przekazania). Będzie tak na~przykład jeśli zadaniem metody \lstinline|setValue| jest ustawienie wartości pola \lstinline|value| (typu \lstinline|Obj|) obiektu na wartość przekazaną jako argument metody \lstinline|setValue| i równocześnie typ \lstinline|Obj| jest właścicielem danych zaalokowanych dynamicznie, powiedzmy:
\begin{lstlisting}
class Obj {
  int* data;  //dane
  int size;   //rozmiar danych
 public:
  Obj(const Obj& src): data(new int[src.size]), size(src.size) {
    //kopiowanie danych:
    std::copy(src.data, src.data+size, data);
  }
  Obj& operator=(const Obj& src) {
    Obj cpy = src;
    this->swap(cpy);
    return *this;
  }
  ~Obj() { delete data; }
  void swap(Obj& o) {
    std::swap(data, o.data);
    std::swap(size, o.size);
  }
  //...inne konstruktory i metody
};
\end{lstlisting}
Proszę zwrócić uwagę, że~powyższy kod (szczególnie operator przypisania) jest odporny zarówno na rzucanie wyjątków przez \lstinline|new|, tj. nie zmienia w~takim wypadku stanu obiektu docelowego, jak i na pseudo-przypisania (typu \lstinline|x=x|). Więcej informacji można znaleźć w~\cite{faqlite}.

W~C++03 metoda \lstinline|setValue| można napisać na jeden z~snastępujących sposobów:
\begin{lstlisting}
void setValue(const Obj& o) { // 1
  this->value = o;
}
void setValue(Obj& o) {       // 2
  this->value = o;
}
void setValue(Obj o) {        // 3
  this->value.swap(o);
} //tu, destruktor obiektu o skasuje stare dane this
\end{lstlisting}

Implementacja (1) nie wie czy~argument jest obiektem tymczasowym, więc na~wszelki wypadek nie może go~zmienić. Jest ona jednak optymalna dla obiektów nietymczasowych.

Wersja (2) nie wnosi nic~pozytywnego w~stosunku do (1), a~dodatkowo nie można jej wywołać na~rzecz obiektów stałych i~tymczasowych (\lstinline|z.setValue(Obj())| nie skompiluje się).

Implementacja (3) w~C++03 może wykonać kopiowanie już przy przekazywaniu parametru do~funkcji, potem zaś dodatkowo wykona metodę \lstinline|swap|.
Pozornie jest ona zatem mniej efektywna niż (1). Jednakże, dla~argumentu będącego \rr-wartością, kompilator może wygenerować kod nie wykonujący wspomnianego kopiowania\footnote{Może to~zrobić nawet gdy konstruktor kopiujący \lstinline|Obj| powoduje efekty uboczne. Dlatego konstruktory kopiujące nigdy nie powinny ich powodować!}, zaś nadmiarowa operacja \lstinline|swap| jest dość ,,tania'' w~stosunku do~kopiowania.

W C++11 można przekazać argument przez referencję do~\rr-wartości:
\begin{lstlisting}
void setValue(Obj&& o) {      // 4
  this->value.swap(o);
} //destruktor obiektu o skasuje stare dane this
\end{lstlisting}
Proszę zwrócić uwagę, że ta~metoda zmienia stan obiektu \lstinline|o|, ale~ponieważ jest on~przekazany przez referencję do~\rr-wartości, to~mamy pewność\footnote{konkretniej: wywołujący składa obietnicę że tak~jest, ale jeśli jej nie dotrzyma to~jego problem}, że, po~powrocie, nie~będzie on~używany aż do~momentu skasowania.

Dobrym rozwiązaniem jest umiejscowienie logiki przenoszącej obiekty klasy \lstinline|Obj| w~przenoszących konstruktorze i~operatorze przypisania tej~klasy:
\begin{lstlisting}
Obj(Obj&& o): data(src.data), size(src.size) {
  o.data = nullptr; //destruktor obiektu o nie skasuje danych
}
Obj& operator=(Obj&& o) {
  this->swap(o);
  return *this;
} //destruktor obiektu o skasuje stare dane this
\end{lstlisting}
Przenoszącą metodę \lstinline|setValue| można teraz zaimplementować czytelniej:
\begin{lstlisting}
void setValue(Obj&& o) {      // 5
  this->value = std::move(o);
  // pisząc std::move obiecujemy,
  // że nie będziemy później, tj. w tym miejscu,
  // używać obiektu o
}
\end{lstlisting}
Użycie \lstinline|std::move| jest wymagane, inaczej zostałby wywołany kopiujący operator przypisania zamiast przenoszącego. Proszę zauważyć, że~wewnątrz metody \lstinline|setValue| obiekt \lstinline|o| żyje aż do jej końca i w~związku z~tym zachowuje się jak typ \lstinline|Obj&|. Potencjalnie, \lstinline|o| mógłby być używany po przypisaniu, dlatego trzeba jawnie go~rzutować na~referencję do~\rr-wartości (to właśnie robi \lstinline|std::move|) i~tym samym pozwolić na~zniszczenie jego stanu.

Ostatecznie, optymalnie jest wyposażyć naszą klasę w~dwie metody \lstinline|setValue|: (1) i~(5).
Alternatywnie, można użyć jedynie wersji (3), która w~przypadku argumentu będącego obiektem tymczasowym, wykona przenoszenie i~operację \lstinline|swap|, czyli także nie spowoduje kosztownego i~niepotrzebnego kopiowania. Przekazywanie argumentów przez wartość będzie miało jeszcze większy sens, gdy tych argumentów będzie więcej, np.
\begin{lstlisting}
void setValues(Obj o1, Obj o2) {
	this->value1.swap(o1);
	this->value2.swap(o2);
}
\end{lstlisting}
jest dobrą alternatywą w~stosunku do~napisania aż 4 przeciążeń o~sygnaturach: \lstinline|void setValues(const Obj& o1, const Obj& o2)|, \lstinline|void setValues(Obj&& o1, Obj&& o2)|, \lstinline|void setValues(const Obj& o1, Obj&& o2)|, \lstinline|void setValues(Obj&& o1, const Obj& o2)|. Ogólnie, dla $n$ argumentów, wystarczy napisać $1$ metodę zamiast $2^n$.

\section{Kopiujące i~przenoszące konstruktory i~operatory przypisania}
Dzięki napisaniu kompletu konstruktorów i~operatorów przypisania w~klasie \lstinline|Obj|, podobny komplet zostanie wygenerowany automatycznie dla takiej klasy:
\begin{lstlisting}
struct ObjHolder {
  Obj obj;
};
\end{lstlisting}
Domyślne kopiujące i~przenoszące konstruktory i~operatory przypisania tej klasy po~prostu odpowiednio kopiują lub przenoszą obiekty pole po polu\footnote{To~wymaga oczywiście, by wszystkie pola miały odpowiedni konstruktor lub operator przypisania. Gdyby np. \lstinline|Obj| nie miał konstruktora kopiującego, to nie~zostałby on~także wygenerowany dla~\lstinline|ObjHolder| (wciąż mogli byśmy jednak taki konstruktor w~\lstinline|ObjHolder| napisać).} (w~przypadku dziedziczenia, wpierw wykonywany jest odpowiedni konstruktor lub~operator z~nad\nobreakdash-klasy, a potem są kopiowane lub przenoszone nowo dodane pola). Gdybyśmy chcieli zaimplementować je~jawnie, wyglądało by~one następująco:
\begin{lstlisting}
struct ObjHolder {
  Obj obj;
  
  //Konstruktory:
  ObjHolder() = default;          //domyślny
  ObjHolder(const ObjHolder& h)   //kopiujący
  	: obj(h.obj) {}
  ObjHolder(ObjHolder&& h)        //przenoszący
  	: obj(std::move(h.obj)) {}
  	
  //operatory przypisania:
  ObjHolder& operator=(const ObjHolder& h) { //kopiujący
    this->obj = h.obj;
    return *this;
  }
  ObjHolder& operator=(ObjHolder&& h) {      //przenoszący
    this->obj = std::move(h.obj);
    return *this;
  }
};
\end{lstlisting}
Proszę też zwrócić uwagę, na~wprowadzoną w~C++11 składnię umożliwiającą przywrócenie konstruktora (w tym przypadku domyślnego).
W analogiczny sposób, jednak za pomocą słowa kluczowego \lstinline|delete|, można skasować wygenerowany konstruktor lub metodę.

\section{Konstruktor przenoszący a zarządzenia zasobami}
Czasami zasadne jest by~klasa posiadała możliwość przenoszenia (przenoszący konstruktor i~operator przypisania) równocześnie nie~umożliwiając kopiowania (np. z~jawnie skasowanymi przenoszącym konstruktorem i~operatorem przypisania).
Taka potrzeba zachodzić najczęściej gdy obiekty takiej klasy są~odpowiedzialne lub~posiadają jakiś zasób który zamykają lub~zwalniają w~destruktorze\footnote{Zgodnie ze wzorcem RAII (Resource Acquisition Is Initialization) łączącym przejęcie i~zwolnienie zasobu z~inicjowaniem i~usuwaniem zmiennych.}. Ponieważ ważne jest by nie~nastąpiło więcej niż jedno wywołanie destruktora zwalniającego zasób, to~konstruktor kopiujący jest niedostępny.
Zaś konstruktor przenoszący, po~prostu przenosi odpowiedzialność za~zasób na~nowy obiekt.
Przykładowo, klasa \lstinline|std::fstream| z~biblioteki standardowej, reprezentująca strumień umożliwiający dostęp do pliku, posiada zasób w~postaci otwartego deskryptora pliku (który jest zamykany przez destruktor). Dlatego \lstinline|std::fstream| nie posiada konstruktora kopiującego, co w~C++03 uniemożliwiło np. zwrócenie obiektu tej klasy jako wyniku funkcji, np.:
\begin{lstlisting}
std::fstream open_config_file() {
	std::fstream result("myapp.cfg");
	//...
	return result;	// albo: return std::move(result);
}	// niepoprawne w C++03, poprawne w C++11
\end{lstlisting}
Powyższa funkcja jest poprawna w~C++11. Plik jest otwierany i~zarządza nim obiekt lokalny \lstinline|result|. Następnie konstruktor przenoszący jest używany by~zwrócić obiekt i~przenieść odpowiedzialność za~otwarty deskryptor pliku na~zewnątrz funkcji.

Uwaga: ponieważ \lstinline|result| jest obiektem lokalnym funkcji, to w~linii z~\lstinline|return| staje się \rr-wartością (za chwile będzie skasowany), nawet gdy nie użyje się \lstinline|std::move| by~jawnie rzutować na~\rr-wartość.

W STL-u istnieją dwa szablony (inteligentne wskaźniki) które ułatwiają zarządzanie zasobami (będących najczęściej w~postaci wskaźników):
\begin{itemize}
	\item \lstinline|std::unique_ptr| --- analogicznie do \lstinline|std::fstream|, zwalnia zasób (typowo: kasuje wskazywany obiekt) w~destruktorze i~jest niekopiowalny;
	\item \lstinline|std::shared_ptr| --- zlicza\footnote{np. w konstruktorze kopiującym zwiększa wartość licznika referencji o 1, zaś w~destruktorze ją o~1 zmniejsza} liczbę referencji do~wskazywanego obiektu i~kasuje go gdy licznik wskaże zero. W ten sposób jeden obiekt jest współdzielony przez kilka wskaźników, zaś operacja ich kopiowania jest dobrze zdefiniowana.
\end{itemize}

\section{Użyj C++11 a Twój program sam przyspieszy}
STL intensywnie korzysta z~nowych możliwości jakie daje C++11 w używaniu \rr-wartości.
Większość szablonów klas (np. kontenerów) nie tylko posiada konstruktor przenoszący, ale także używa przenoszenia zamiast kopiowania tam gdzie to~tylko możliwe.
Przykładowo \lstinline|std::vector| przenosi (zamiast, jak było to w~C++03, kopiować) przechowywane obiekty w~trakcie, dokonywanej od~czasu do~czasu, realokacji, która jest wymagana by dostosować ilość używanej pamięci do~liczby przechowywanych obiektów.

Napisany w~C++03 program, często, by działał szybciej, wystarczy jedynie skompilować za~pomocą C++11.

\bibliographystyle{plplain}
\nocite{*}
\bibliography{cpp}

\begin{thebibliography}{1}

\bibitem{faqlite}
C++ faq lite.
\newblock \url{http://klub.chip.pl/b.krzemien/c++-faq-pl/index.html}.
\newblock [dostęp: 2013-04-27].

\bibitem{cppref}
C++ reference.
\newblock \url{http://en.cppreference.com}.
\newblock [dostęp: 2013-04-27].

\bibitem{enwiki:cpp11}
Wikipedia, the free encyclopedia: C++11.
\newblock \url{http://en.wikipedia.org/wiki/C++11}.
\newblock [dostęp: 2013-04-27].

\bibitem{plwiki:cpp11}
Wikipedia, wolna encyklopedia: C++11.
\newblock \url{http://pl.wikipedia.org/wiki/C++11}.
\newblock [dostęp: 2013-04-27].

\bibitem{CPP11STD}
ISO.
\newblock {\em ISO/IEC N3337=12-0027 Working Draft, Standard for Programming
  Language C++}.
\newblock International Organization for Standardization, 2012.

\end{thebibliography}

\end{document}